\documentclass[aps,prb,twocolumn,superscriptaddress,longbibliography]{revtex4-2}
\usepackage{epsfig,amsopn}
\usepackage{graphicx}
\usepackage{color}
\usepackage{amsmath,amssymb}
\usepackage{enumerate}
\usepackage{verbatim}
\usepackage{braket}
\newcommand\bea{\begin{eqnarray}}
\newcommand\eea{\end{eqnarray}}
\newcommand\beq{\begin{equation}}
\newcommand\eeq{\end{equation}}

\def\nn{\nonumber}
\def\f{\frac}
\def\al{\alpha}

\def\Do{\partial}

\def\ua{\uparrow}
\def\da{\downarrow}

\def\be{\beta}

\def\th{\theta}

\def\inf{\infty}

\begin{document}
\title{Orientation dependent anomalous Hall and spin Hall currents at junctions of altermagnets with $p$-wave magnets} 
\author{Sachchidanand Das}
\affiliation{School of Physics, University of Hyderabad, Prof. C. R. Rao Road, Gachibowli, Hyderabad-500046, India}

 \author{ Abhiram Soori}
 \email{abhirams@uohyd.ac.in}
 \affiliation{School of Physics, University of Hyderabad, Prof. C. R. Rao Road, Gachibowli, Hyderabad-500046, India}
\begin{abstract}
We study charge and spin transport across a junction between an altermagnet (AM) and a $p$-wave magnet (PM) using a continuum model with boundary conditions tailored to the spin-split band structures of the two materials. Remarkably, although neither AM nor PM is spin-polarized, we find that the junction supports finite spin currents both longitudinally and transversely. We compute the longitudinal and transverse charge and spin conductivities as functions of the crystallographic orientations and the relative angle between the N\'eel vectors of AM and PM. Our results reveal that transverse charge and spin conductivities can be finite even when the longitudinal charge conductivity vanishes. For suitable parameter choices and orientation angles, the transverse conductivities are more prominent than the longitudinal ones. The origin of these effects lies in the matching and mismatching of transverse momentum modes ($k_y$) across the junction combined with the spin-dependent band splitting in AM and PM. Furthermore, while the transverse charge conductivity may be zero for certain orientations, the transverse spin conductivity remains finite due to unequal contributions of opposite $k_y$ channels. These findings highlight AM–PM junctions as a promising platform for tunable generation and control of transverse charge and spin currents driven purely by crystallographic orientation and spin structure.
\end{abstract}
\maketitle
\section{Introduction}
In recent years, AMs ~\cite{smejkal22a} have emerged as a highly popular research topic in condensed matter physics due to their unique and unconventional properties. What makes them particularly fascinating is that they combine features of both ferromagnets and antiferromagnets (AFM). Like AFM, AMs exhibit zero net magnetization, meaning they do not produce an external magnetic field. However, unlike typical AFM, they display spin-split electronic band structures and time reversal symmetry breaking similar to those found in ferromagnets~\cite{smejkal20,feng22}. This spin-splitting usually requires a net magnetic moment, but in AMs, it arises purely from the bandstructure  of the crystal lattice. As a result, AMs can provide spin-polarized electrons without generating stray magnetic fields, which is a major advantage for spintronic applications. When a bias voltage is applied across a junction between a normal metal and an AM, a spin current is generated--demonstrating their potential in spin-based electronic devices \cite{Das23}.  

AMs  can be viewed as the magnetic counterparts of d-wave superconductors. In this analogy, s-wave superconductors correspond to ferromagnets, while $p$-wave superconductors resemble spin-orbit-coupled systems. The magnetic equivalent of the anisotropic triplet pairing is known as PMs. Much like AMs, PMs also display spin-split electronic band structures. However, a key difference is that, unlike d-wave AMs, PMs preserve time-reversal symmetry~\cite{Hellenes24}. While they share several qualitative features with spin-orbit-coupled systems, the band structure of PMs is notably anisotropic, especially around  $\vec{k}=0$. Recent studies on PMs show that these magnets can even coexist with superconductivity and enable strong charge-to-spin conversion and transverse spin current~\cite{Sukhachov25}. Research on  the junction with PMs found significant magnetoresistance and spin-filtering effects along with anisotropic bulk spin conductivity~\cite{Brekke24}. Junctions of normal metals with PMs  under applied bias generate transverse spin currents~\cite{Hedayati24}. The Pauli spin matrix (or a linear combination of Pauli spin matrices) that commutes with the Hamiltonian for AM/PM defines the N\'eel vector for AM/PM. 

The spin Hall effect in spin-orbit-coupled metals, where opposite spins accumulate at opposite edges of the system in the transverse direction, has been widely investigated~\cite{murakami03,sinova04,shermp}. Giant magnetoresistance, a phenomenon that has found widespread application in modern data storage technologies~\cite{chappert07}, has also been extensively studied in ferromagnets. The combination of ferromagnets with materials possessing strong spin–orbit coupling has further enabled the development of devices such as the Datta–Das spin transistor~\cite{dattadas,chuang2015,Choi2015,bijay22}. More recently, antiferromagnetic spintronics has emerged as a rapidly growing area of research, expanding the range of spin-related effects accessible in experiments~\cite{baltz18,fukami20,hoffmann22}. 

Transverse currents can arise in response to a longitudinal bias when a perpendicular magnetic field is applied, an effect known as the Hall effect. In spin-orbit-coupled metals, the application of a Zeeman field similarly induces transverse currents~\cite{soori2021}. In ferromagnetic materials with spin–orbit coupling, transverse currents can  appear even in the absence of an external magnetic field, giving rise to the anomalous Hall effect~\cite{nagaosa2010}.

The spin Hall effect in AMs does not rely on spin–orbit coupling, as demonstrated in recent studies~\cite{bai22,karube22,bose22}. Hall responses in AMs can also be generated optically~\cite{Farajollahpour2025}. Moreover, transport across hybrid heterostructures involving AMs or PMs coupled to superconductors has been actively explored~\cite{Papaj23,sun23,Das24,Soori25,Nagae25,zeng25,Fukaya25}. Also, strong dependence on the orientation of the crystallographic axis of AMs in Josephson junctions with AMs at the center has been studied by various groups~\cite{Ouassou23,Cheng24B,Cheng24A,Beenakker23}. In parallel, non-Hermitian extensions of AM and PM models have opened new avenues for investigating exotic transport and spectral properties~\cite{Reja24,reja25,alipourzadeh25}. These developments highlight the growing recognition of AMs as a fertile platform for spintronic functionalities beyond conventional spin–orbit-coupled systems.

Motivated by this progress, we study charge and spin transport across a junction of an AM and a PM within a continuum framework. We consider arbitrary rotations of the crystallographic axes of both materials, and employ transverse momentum matching—made possible by translational invariance along the transverse direction—to compute the conductivities on either side of the junction under a longitudinal bias. Our analysis reveals the coexistence of longitudinal charge current with both transverse charge and transverse spin currents, thereby demonstrating that anomalous Hall and spin Hall effects can arise in AM–PM junctions even in the absence of spin–orbit coupling. In addition, we find that both materials support finite spin currents despite not being spin-polarized. By varying the N\'eel vector directions of the AM and PM, we further uncover how their relative alignment influences the transport characteristics. These results establish AM–PM junctions as a new platform where unconventional spin and Hall responses emerge without relying on spin–orbit coupling.

\begin{figure} [htb]
    \centering
    \includegraphics[width=9.0cm,height=4cm]{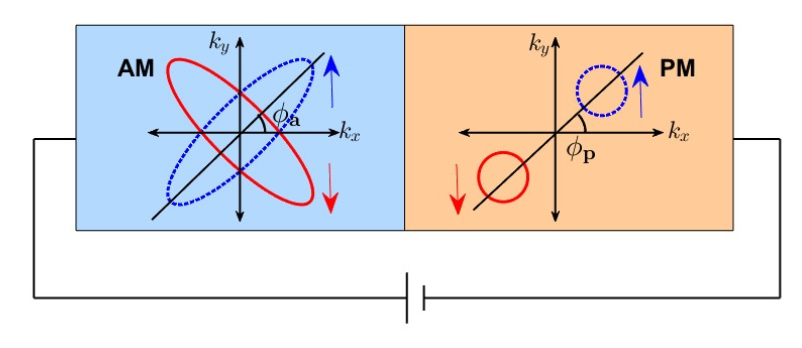} 
    \caption{Schematic of the AM–PM junction: light blue rectangle on the left shows AM whereas orange rectangle on the right shows PM. AM is rotated by angle $\phi_a$ and  PM rotated by angle $\phi_p$. Fermi surfaces (FS): blue solid loops are for up-spin and red dashed loops are for down-spin. Elliptical loops on the left represent FS's of AM and circular loops on the right represent FS's of PM. }
    \label{fig:schm}
\end{figure}

In section~\ref{sec:calc}, we show details of the calculation. In section~\ref{sec:result}, we show obtained results on longitudinal/transverse, charge/spin conductivities for realistic choice of parameters, along with reasoning. In section~\ref{sec:disc}, we  compare our results with those in other platforms and discuss the advantage of this method over the lattice model.  Finally in section~\ref{sec:summary}, we summarize our work and conclude. 

\section{Calculations}~\label{sec:calc}

We consider a system consisting of an AM on the left extending from $-\inf<x<0$  and  a PM on the right ranging from $0<x<\inf$, and  both ranging from $-\inf<y<\inf$ along the transverse direction making a junction at $x=0$. Since this system is translationally invariant along $y$ direction, so the momentum associated with the transverse direction is conserved and we can study the electron transport in such a system by transverse momentum matching on both the sides. Hamiltonian for such a system  is given by

\bea 
H &=& -\Big[t_0\sigma_0-t_J\sigma_z\cos2\phi_a\Big]a^2\Do_x^2 - \Big[t_0\sigma_0 + \nn \\  && t_J\sigma_z\cos2\phi_a\Big]a^2\Do_y^2 +2t_J\sigma_za^2\Do_x\Do_y~\sin2\phi_a \nn \\  && - \mu_{a} ~~~~~~~~~~~~~~~~~~~~~~~~~~~~{\rm for ~~}~~ x<0  \nn \\
  &=& \Big[-ta^2\vec{\nabla}^2-\mu_p\Big]\sigma_0-i\al a (\hat{n}_{\phi_p}.\vec{\nabla})~\hat{n}_{\beta}.\vec{\sigma},\nn \\ &&~~~~~~~~~~~~~~~~~~~~~~~~~~~~~~~~~~~~~~~~~{\rm for ~~} x>0 
  .\label{eq:Ham-cont} 
\eea

where $t_0$ is the hopping strength , $t_J$ is the spin and direction dependent hopping, and $\mu_{a}$ is the chemical potential in the AM. $\phi_a$ is the angle by which crystallographic plane of the AM is rotated and $a$ is the lattice constant. $t$ is the hopping strength  in PM (we take $t=t_0$ for simplicity), $\alpha$ is the strength of the term that characterizes PM and $\hat{n}_{\phi_p}=\cos{\phi_p}~\hat{x} + \sin{\phi_p}~\hat{y}$, $\phi_p$ being  the angle of crystallographic orientation with respect to $x$-axis in anti-clockwise direction, $\hat{n}_{\beta}=\cos{\beta}~\hat{z} + \sin{\beta}~\hat{x}$ ,  $\beta$ is the angle between the spin quantization axes of AM and PM, $\vec{\sigma}=\sigma_x\hat{x}+\sigma_y\hat{y}+\sigma_z\hat{z}$ where  $\sigma_x,\sigma_y, \sigma_z$ are the Pauli spin matrices and $\mu_p$ is the chemical potential in the PM. Eigenvalues of this matrix are $\pm1$ with eigenvector $\ket{\ua_{\beta}}=[\cos{\beta/2},~\sin{\beta/2}]^T$corresponding to +1 and  $\ket{\da_{\beta}}=[-\sin{\beta/2},~\cos{\beta/2}]^T$ corresponding to -1.
 
 Dispersion relation for the PM is given by 
\beq
 E_{PM} = \Big[t(k_x^2a^2+k_y^2a^2)-\mu_p\Big] + \eta\alpha (\cos{\phi_p}~k_xa + \sin{\phi_p}~k_ya)~~ \\
\eeq
where $\eta=1$ for up-spin electrons, $\eta=-1$ for down-spin electrons with respect to $\hat{n}_{\beta}$.\\
Dispersion for AM with up spin  and down spin electrons are given by
\begin{align}
    E_{AM} &= \Big[t_0\sigma_0-t_J\eta\cos2\phi_a\Big]k_x^2a^2 + \Big[t_0\sigma_0  + t_J\eta\cos2\phi_a\Big] k_y^2a^2 \nn \\ &-2t_J\eta k_xk_ya^2~\sin2\phi_a   - \mu_{a} 
\end{align} 

The expressions for   longitudinal and transverse charge current  densities on the AM  are given by 
\begin{align}
    J_{x,am}=\frac{2ea^2}{\hbar}{\rm Im}\big[\psi^{\dagger}\big(t_0\sigma_0-t_J\cos2\phi_a\sigma_z)\Do_x\psi \nn \\
    - it_Jk_y\sin2\phi_a\psi^{\dagger}\sigma_z\psi\Big]
    \label{eq:Jxam}
\end{align}  
  and 
\begin{align}
    J_{y,am}=\frac{2ea^2}{\hbar}\Big[t_0k_y\psi^{\dagger}\psi+t_Jk_y\cos2\phi_a\psi^{\dagger}\sigma_z\psi \nn \\ 
    -t_J\sin2\phi_a{\rm{Im}}(\psi^{\dagger}\sigma_z\Do_x\psi)\Big] \label{eq:Jyam}
\end{align}
 respectively,  whereas the longitudinal and transverse  spin current  densities for the same are given by
 \begin{align}
     J^s_{x,am}={a^2\rm{Im}}\big[\psi^{\dagger}\big(t_0\sigma_z-t_J\cos2\phi_a\sigma_0)\Do_x\psi \nn \\ 
     - it_Jk_y\sin2\phi_a\psi^{\dagger}\sigma_0\psi\Big]
     \label{eq:Jxams}
 \end{align} 
 and \begin{align}
     J^s_{y,am}=a^2\Big[t_0k_y\psi^{\dagger}\sigma_z\psi+t_Jk_y\cos2\phi_a\psi^{\dagger}\sigma_0\psi \nn \\ 
     -t_J\sin2\phi_a{\rm{Im}}(\psi^{\dagger}\sigma_0\Do_x\psi)\Big] 
     \label{eq:Jyams}
 \end{align}  respectively.
 
Similarly, the transverse and longitudinal charge current densities in  the PM are given by 
\begin{align}
    J_{x,pm}=\frac{a^2e}{\hbar}\Big[{2~\rm{Im}}\Big(t\psi^\dagger\sigma_0\Do_x\psi\Big)+\frac{\al}{a}\cos\phi_p\psi^{\dagger}\sigma_\beta\psi\Big]
    \label{eq:Jxpm}
\end{align}  
 and 

\begin{align}
    J_{y,pm}=\frac{a^2e}{\hbar}\Big[{2~\rm{Im}}\Big(t\psi^\dagger\sigma_0\Do_y\psi\Big)+\frac{\al}{a}\sin\phi_p\psi^{\dagger}\sigma_\beta\psi\Big]
    \label{eq:Jypm}
\end{align}  respectively, whereas the transverse and longitudinal spin current densities in the PM are given by 
\begin{align}
    J^s_{x,pm}={a^2\rm{Im}}\Big(t\psi^\dagger\sigma_\beta\Do_x\psi\Big)+\frac{\alpha a}{2}\cos\phi_p\psi^{\dagger}\sigma_0\psi
    \label{eq:Jxpms} 
\end{align} and \begin{align}
    J^s_{y,pm}={a^2\rm{Im}}\Big(t\psi^\dagger\sigma_\beta\Do_y\psi\Big)+\frac{\alpha a}{2}\sin\phi_p\psi^{\dagger}\sigma_0\psi
    \label{eq:Jypms}
\end{align} 

By the conservation of longitudinal charge current at $x=0$ we find the boundary conditions which are given below
\begin{align}
     \psi_L &= c~\psi_R,~~ \nn \\ 
    &c\Big[(t_0\sigma_0-t_J\sigma_z\cos2\phi_a)a\Do_x-it_J\sigma_z\sin2\phi_a k_ya\Big] \psi_L \nn \\
      &= \Big[ta\Do_x\sigma_0 +\big(\frac{i\alpha}{2}\cos{\phi_p}\sigma_{\beta}+V_0\sigma_0\big)\Big]\psi_R, \label{eq:bc}
\end{align}
where $\psi_L=\psi(0^-)$, $\Do_x\psi_L=\Do_x\psi|_{0^-}$, $\psi_R=\psi(0^+)$ and $\Do_x\psi_R=\Do_x\psi|_{0^+}$.

We write down the scattering eigenfunctions and the charge/spin, longitudinal/transverse currents in different regions for an up-spin electron incident at the junction in subsection~\ref{sec:up-spin}, followed by that for down spin incidence in subsection~\ref{sec:down-spin}.  Using the expressions for different currents we outline a method to calculate the conductivities in subsection~\ref{sec:conduct}.  Then we calculate scattering coefficients analytically in some particular limit in subsection~\ref{sec:beta} which is useful later for explaining the behavior of charge and spin conductivities versus $V_0$. 

  \subsection{Up-spin incidence}~\label{sec:up-spin}
  For this case, the Pauli matrices do not commute with the Hamiltonian. Hence, if we incident an up-spin electron from the AM  side, then the same electron can reflect back  either with the same spin or with the opposite spin. Similarly, the electron can also get transmitted to the $p$-wave region either as $\ket{\ua_{\beta}}$ or as $\ket{\da_{\beta}}$. When an electron with up-spin is incident at the AM/$p$-wave junction  with energy $E$ making an angle $\theta$, the wavefunction is given by  $\psi(x)e^{ik_{y,\ua}y}$, where
  \begin{align}
      \psi(x) &= \Big(e^{ik_{r,\ua} x}~\ket{\ua}+r_{\ua\ua}e^{ik_{l,\ua} x}~\ket{\ua}+  \nn  \\
&~~~~~~r_{\da\ua}e^{ik_{l,\da}x}~\ket{\da}\Big)~~~~~~~~~~~{\rm for ~~}x<0~~ \nn \\
~~~~&=\Big(t_{\ua\ua}e^{ik'_{x,\ua} x}~\ket{\ua_{\beta}}+t_{\da\ua}e^{ik'_{x,\da}x}~\ket{\da_{\beta}}\Big) \nn \\
&~~~~~~~~~~~~~~~~~~~~~~~~~~~~~~~~~~~~~~~~{\rm for ~~}x>0 \label{eq:psi1}
  \end{align}

where $r_{\ua\ua}$ and $r_{\da\ua}$ are the reflection amplitude for $\ua$ and $\da$ electrons respectively whereas $t_{\ua\ua}$ and $t_{\da \ua}$ are the transmission amplitude for $\ket{\ua_{\beta}}$ and $\ket{\da_{\beta}}$. $\ket{\ua}=[1~~0]^T$ and  $\ket{\ua}=[0~~1]^T$. The system is translationally invariant along transverse direction, making $k_{y,\ua}$  a good quantum number.  From the dispersion, $k_x$ can be obtained once $k_y$ is known. We calculate the group velocity $v_g={dE}/{\hbar dk_x}$ and decide which $k_x$ is left-mover and which is right-mover. The wavenumbers that enter the wavefunction in Eq.~\ref{eq:psi1} are given by: 
\beq
k_{r,\ua}a = \sqrt{\frac{(E+\mu_a)}{(t_0-t_J)}}~\cos{\th}\cos\phi_a+\sqrt{\frac{E+\mu_a}{t_0+t_J}}~\sin{\th}\sin\phi_a \nn
\eeq

\beq
k_{l,\ua}a= \sqrt{\frac{(E+\mu_a)}{(t_0-t_J)}}~\cos{\th_l}\cos\phi_a+\sqrt{\frac{E+\mu_a}{t_0+t_J}}~\sin{\th_l}\sin\phi_a \nn
\eeq

\beq
k_{y,\ua}a = \sqrt{\frac{(E+\mu_a)}{(t_0-t_J)}}~\cos{\th}\sin\phi_a+\sqrt{\frac{E+\mu_a}{t_0+t_J}}~\sin{\th}\cos\phi_a \nn
\eeq 
 $k_{r,\ua}$ is the wavevector of the particles moving towards right, possessing  positive velocity whereas $k_{l,\ua}$ and $k_{l,\da}$ are the wavevector of the up-spin and down-spin electrons respectively which get reflected from the boundary $x=0$ and possess a negative velocity. $k_{l,\da}$ is calculated by the down spin dispersion in the AM.  Only those values of $\th$ are allowed for which $k_{r/l,\ua}$ is real. So the range of $\th$ is given by $\th \in \Big(-\frac{\pi}{2}-\eta, \frac{\pi}{2}-\eta\Big) $. $\th_l$ is chosen such that, for $k_{l,\ua}$ the $v_g$ of the left moving electron is negative. So,
$\th_l=\th-\pi-2\eta$ where  $\eta=\tan^{-1}\Big(\sqrt{{(t_0+t_J)/(t_0-t_J)}}\tan\phi_a\Big)$.
$k'_{x,\ua}$ is the wavevector of up-spin transmitted electrons in the PM whereas $k'_{x,\da}$ is the wavevector for down-spin transmitted electrons and they are chosen such that either the wave is a forward mover or it decays in the PM. The expressions for the two are given below--

\begin{align}
    k'_{x,\ua}a &= -\frac{\al\cos\phi_p}{2t} ~\pm\nn \\ & \frac{1}{2}\sqrt{\frac{\al^2\cos^2\phi_p}{t^2}-4\Bigg(k_{y,\ua}^2a^2+\frac{\al\sin\phi_p}{t}k_{y,\ua}a-\frac{E+\mu_p}{t}\Bigg)}\nn
\end{align}

\begin{align}
    k'_{x,\da}a &= \frac{\al\cos\phi_p}{2t} ~\pm\nn \\ & \frac{1}{2}\sqrt{\frac{\al^2\cos^2\phi_p}{t^2}-4\Bigg(k_{y,\ua}^2a^2-\frac{\al\sin\phi_p}{t}k_{y,\ua}a-\frac{E+\mu_p}{t}\Bigg)}\nn
\end{align}


\subsection{Down-spin incidence}~\label{sec:down-spin}
When a down spin electron with energy $E$ is incident from the AM to the AM/$p$-wave junction at an angle $\theta$, it may get reflected as the same down spin or as an up-spin electron in the AM and get transmitted to the PM either as $\ket{\da_{\beta}}$ or as $\ket{\ua_{\beta}}$. The wavefunction for such a process is given by $\psi(x)e^{ik_{y,\da}y}$, 
where
  \begin{align}
      \psi(x) &= \Big(e^{ik_{r,\da} x}~\ket{\da}+r_{\da\da}e^{ik_{l,\da} x}~\ket{\da}+  \nn  \\
&~~~~~~r_{\ua\da}e^{ik_{l,\ua}x}~\ket{\ua}\Big)~~~~~~~~~~~{\rm for ~}~x<0~~ \nn \\
~~~~&=\Big(t_{\ua\da}e^{ik'_{x,\ua} x}~\ket{\ua_{\beta}}+t_{\da\da}e^{ik'_{x,\da}x}~\ket{\da_{\beta}}\Big) \nn \\
&~~~~~~~~~~~~~~~~~~~~~~~~~~~~~~~~~~~~~~~~{\rm for ~~}x>0 \label{eq:psi2}
  \end{align}

where $r_{\da\da}$ and $r_{\ua\da}$ are the reflection amplitude for $\da$ and $\ua$ respectively whereas $t_{\ua\da}$ and $t_{\da \da}$ are the transmission amplitude for $\ket{\ua_{\beta}}$ and $\ket{\da_{\beta}}$ respectively. $k_{r,\da}$ is the wavevector of the down-spin incident electron whereas $k_{l,\da}$  is the wavevector of the same spin reflected electron at the boundary. $k_{y,\da}$ is the transverse wavevector of the down spin electron. Different wavevectors that enter the expression for the wavefunction in Eq.~\ref{eq:psi2} are given by 
  \beq
k_{r,\da}a = \sqrt{\frac{(E+\mu_a)}{(t_0+t_J)}}~\cos{\th}\cos\phi_a+\sqrt{\frac{E+\mu_a}{t_0-t_J}}~\sin{\th}\sin\phi_a \nn
\eeq

\beq
k_{l,\da}a= \sqrt{\frac{(E+\mu_a)}{(t_0+t_J)}}~\cos{\th_{l}}\cos\phi_a+\sqrt{\frac{E+\mu_a}{t_0-t_J}}~\sin{\th_{l}}\sin\phi_a \nn
\eeq

\beq
k_{y,\da}a = \sqrt{\frac{(E+\mu_a)}{(t_0+t_J)}}~\cos{\th}\sin\phi_a+\sqrt{\frac{E+\mu_a}{t_0-t_J}}~\sin{\th}\cos\phi_a, \nn
\eeq 
where  $\th$ is the angle of incidence and  only those values of $\th$ are allowed for which $k_{r/l,\da}$ is real. The range of $\th$ is given by $\th \in \Big(-\frac{\pi}{2}-\eta_2, \frac{\pi}{2}-\eta_2\Big) $. $\th_l$ is chosen such that, for $k_{l,\da}$ the $v_g$ of the left moving electron is negative. So,
$\th_{l}=\th-\pi-2\eta_2$ where  $\eta_2=\tan^{-1}\Big(\sqrt{{(t_0-t_J)/(t_0+t_J)}}\tan\phi_a\Big)$.
The interpretation of $k'_{x,\ua}$ and $k'_{x,\da}$ are the same as described earlier. They are  the wavevectors of up- and down-spin transmitted electrons respectively in the PM, but now for down-spin electron incidence. The expressions for the two are given below
\begin{align}
    k'_{x,\ua}a &= -\frac{\al\cos\phi_p}{2t} ~\pm\nn \\ & \frac{1}{2}\sqrt{\frac{\al^2\cos^2\phi_p}{t^2}-4\Bigg(k_{y,\da}^2a^2+\frac{\al\sin\phi_p}{t}k_{y,\da}a-\frac{E+\mu_p}{t}\Bigg)}\nn
\end{align}
\begin{align}
    k'_{x,\da}a &= \frac{\al\cos\phi_p}{2t} ~\pm\nn \\ & \frac{1}{2}\sqrt{\frac{\al^2\cos^2\phi_p}{t^2}-4\Bigg(k_{y,\da}^2a^2-\frac{\al\sin\phi_p}{t}k_{y,\da} a-\frac{E+\mu_p}{t}\Bigg)}\nn
\end{align}
One of the  $\pm$ signs is chosen here such that either the wave is forward mover or it    decays in the PM.  $k_{l,\ua}$ is determined using the up-spin dispersion, taking $k_y=k_{y,\da}$. 
\subsection{Conductivities}~\label{sec:conduct}
The scattering coefficients are obtained from the boundary conditions in Eq.~\ref{eq:bc}. Once these coefficients are determined, the wavefunctions can be constructed, from which the charge and spin current densities are evaluated. These current densities form the basis for calculating the longitudinal and transverse charge and spin conductivities on both sides of the junction.  

The longitudinal conductivities are obtained from the longitudinal charge and spin current densities as   
\begin{align}
 G= \frac{e}{8\pi^2a^2}\Bigg[\frac{1}{\sqrt{t_0^2-t_J^2}}\Bigg(\int{J_{x}^\ua~d\theta+\int J_{x}^\da}d\theta\Bigg)\Bigg] \\
    G_{w,s}= \frac{e}{16\pi^2a^2}\Bigg[\frac{1}{\sqrt{t_0^2-t_J^2}}\bigg(\int{J_{x,w}^{s,\ua}d\theta+\int J_{x,w}^{s,\da}}d\theta\bigg)\Bigg] 
 \end{align}
 where $w={\rm AM/PM}$. Here $J_{x}^\ua$ and $J_{x}^\da$ denote the longitudinal charge current densities for up- and down-spin incidence, respectively, and are evaluated using Eq.~\ref{eq:Jxam} in the AM region and Eq.~\ref{eq:Jxpm} in the PM region. Since the longitudinal current is conserved across the junction, the currents in the AM and PM regions are equal. Likewise, $J_{x,w}^{s,\ua}$ and $J_{x,w}^{s,\da}$ denote the longitudinal spin current densities for up- and down-spin incidence, evaluated using Eq.~\ref{eq:Jxams} in the AM region and Eq.~\ref{eq:Jxpms} in the PM region.

 The transverse conductivities on both sides of the junction are obtained from the corresponding transverse charge and spin current densities as   
\begin{align}
    G_{t,w}= \frac{e}{8\pi^2a^2}\Bigg[\frac{1}{\sqrt{t_0^2-t_J^2}}\bigg(\int{J_{y,w}^\ua d\theta+\int J_{y,w}^\da}d\theta\bigg)\Bigg] \\
    G_{t,w}^{s}= \frac{e}{16\pi^2a^2 }\Bigg[\frac{1}{\sqrt{t_0^2-t_J^2}}\bigg(\int{J_{y,w}^{s,\ua}d\theta+\int J_{y,w}^{s,\da}}d\theta\bigg)\Bigg]    
\end{align}
 where $J_{y,w}^\ua$ and $J_{y,w}^\da$ are the transverse charge current densities for up- and down-spin incidence, calculated using Eq.~\ref{eq:Jyam} in the AM region and Eq.~\ref{eq:Jypm} in the PM region. Similarly, $J_{y,w}^{s,\ua}$ and $J_{y,w}^{s,\da}$ denote the transverse spin current densities for up- and down-spin incidence, evaluated using Eq.~\ref{eq:Jyams} in the AM region and Eq.~\ref{eq:Jypms} in the PM region.  

\subsection{Scattering coefficients in the limit $\beta = 0$}~\label{sec:beta}

For simplicity, we set $\phi_a = 0$ and $\phi_p = \pi/2$. In the limit $\beta = 0$, $\sigma_z$ becomes a good quantum number, and the problem decouples into two independent spin sectors. For an incident spin-up electron, the reflection and transmission amplitudes into spin-down states, $r_{\downarrow\uparrow}$ and $t_{\downarrow\uparrow}$, vanish. Likewise, for an incident spin-down electron, the amplitudes $r_{\uparrow\downarrow}$ and $t_{\uparrow\downarrow}$ are zero. Because $\phi_a = 0$, the wavevectors satisfy $k_{r,\uparrow} = -k_{l,\uparrow}$ and $k_{r,\downarrow} = -k_{l,\downarrow}$. These simplifications allow us to obtain exact analytical expressions for the scattering coefficients, which are given by
 \begin{align}
  r_{\ua\ua}=\f{a(t_0-t_J)ik_{r,\ua}-(ik'_{x,\ua}at+V_0)}{a(t_0-t_J)ik_{r,\ua}+(ik'_{x,\ua}at+V_0)} \nn   \\ 
   t_{\ua\ua}=\f{2a(t_0-t_J)ik_{r,\ua}}{a(t_0-t_J)ik_{r,\ua}+(ik'_{x,\ua}at+V_0)} ~\label{eq:V0_up}
 \end{align}
 \begin{align}
  r_{\da\da}=\f{a(t_0+t_J)ik_{r,\da}-(ik'_{x,\da}at+V_0)}{a(t_0+t_J)ik_{r,\da}+(ik'_{x,\da}at+V_0)} \nn \\ 
  t_{\da\da}=\f{2a(t_0+t_J)ik_{r,\da}}{a(t_0+t_J)ik_{r,\da}+(ik'_{x,\da}at+V_0)} ~\label{eq:V0_down}
\end{align}

\section{Results}~\label{sec:result}
The longitudinal charge current is conserved and remains the same in both regions. We therefore evaluate the longitudinal charge conductivity on the PM side. In contrast, the spin current is not conserved (for $\beta \neq 0,\pi$), since no spin component commutes with the full Hamiltonian. On the PM side the spin current corresponds to $\sigma_{\beta}$, while on the AM side it corresponds to $\sigma_{z}$. Unlike the longitudinal current, the transverse charge and spin currents vary with position $x$. We compute the transverse charge and spin conductivities near the junction at $x=0$, on both sides. Now,  we discuss the results on charge and spin conductivities for the PM in subsection~\ref{sec:PM-result} and for the AM in subsection~\ref{sec:AM-result}. 

\subsection{In $p$-wave magnet}~\label{sec:PM-result}

\begin{figure} [htb]
    \centering
    \includegraphics[width=4.25cm,height=3.7cm]{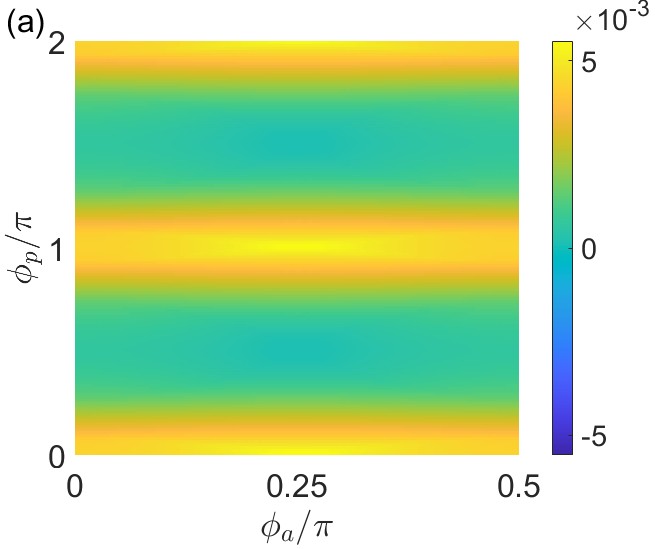}
    \includegraphics[width=4.25cm,height=3.7cm]{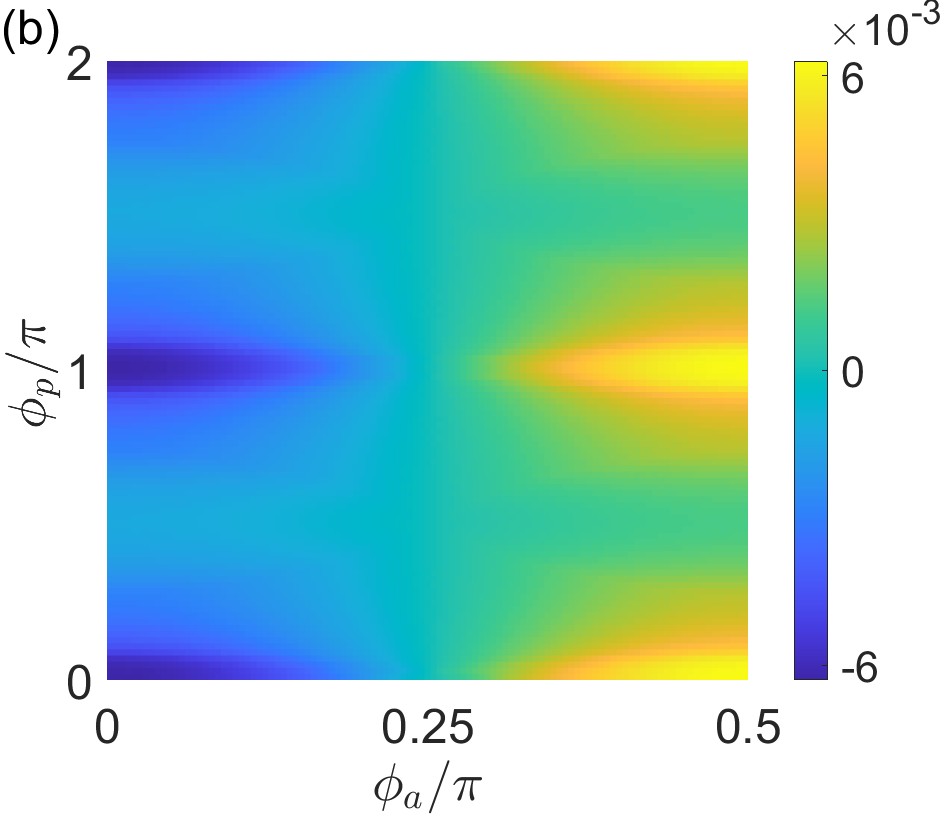} 
   \includegraphics[width=4.25cm,height=3.3cm]{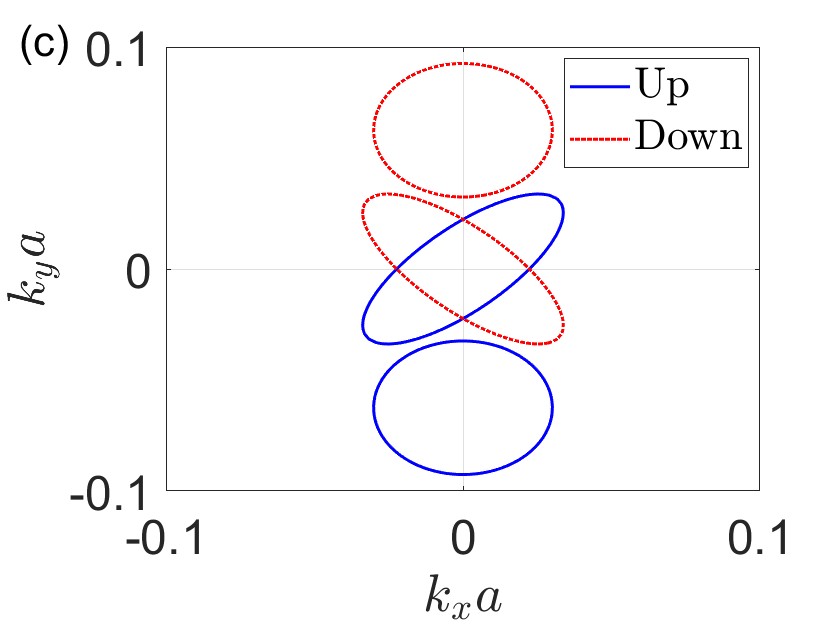}
    \includegraphics[width=4.25cm,height=3.3cm]{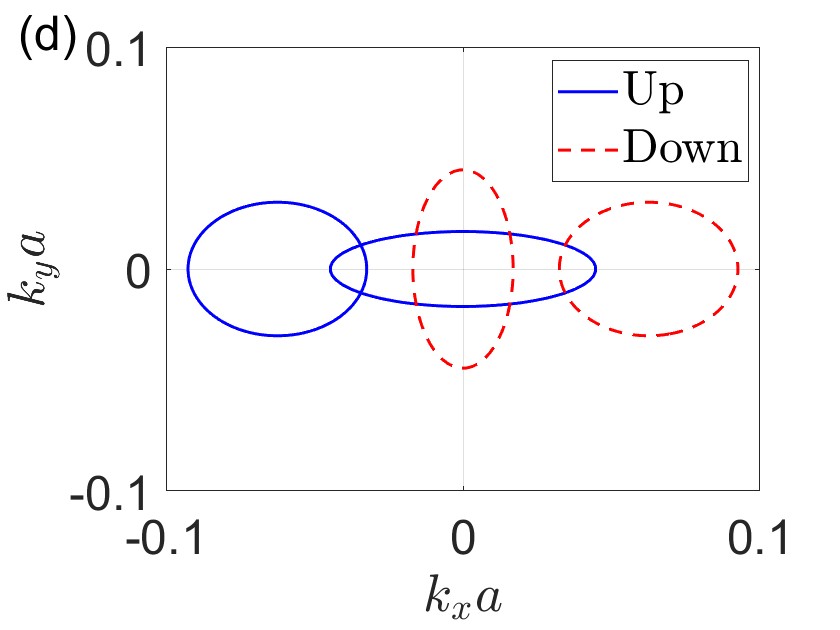} 
    \caption{ (a)~Longitudinal charge conductivity in units of $e^2/ha$ and (b)~longitudinal spin conductivity in units of $e/a$ on the PM versus $\phi_a$ and $\phi_p$, (c,d) Fermi surfaces on the two sides of the junction  for (c) $\phi_a=\pi/4$ and $\phi_p=\pi/2$ and (d) $\phi_a=0$ and $\phi_p=0$ at zero bias for the parameters $t_0=0.1t,~t_J=0.75t_0,~c=1,~\beta=0~,V_0=0,~\alpha=0.125t,~\mu_a=5\times 10^{-4}t_0$ and $\mu_p=-3\times 10^{-3}t$. } 
    \label{fig:G}
\end{figure}

Figure~\ref{fig:G}(a) presents the longitudinal charge  conductivity as a function of the crystallographic rotation angles $\phi_a$ (for the AM) and $\phi_p$ (for the PM) at zero bias. The parameters used in the calculation are $t_0 = 0.1t$, $t_J = 0.75t_0$, $c = 1$, $\beta = 0$, $V_0 = 0$, $\alpha = 0.125t$, $\mu_a = 5\times10^{-4}t_0$, and $\mu_p = -3\times10^{-3}t$. These values are chosen to correspond to realistic material parameters for the altermagnet KRu$_4$O$_8$ and the $p$-wave magnet CeNiAsO.
Figure~\ref{fig:G}(a) exhibits symmetry about $\phi_p=\pi$, since under the transformation $\phi_p \to \phi_p+\pi$, the Fermi surfaces of the two spin species in the PM interchange. In the AM, the Fermi surfaces satisfy the property that if $(k_x,k_y)$ lies on the Fermi surface, so does $(-k_x,-k_y)$ for both spin species. Together with the transverse momentum matching condition at the interface, these features result in the observed symmetry about $\phi_p=\pi$. The longitudinal conductivity nearly vanishes at $\phi_a=\pi/4$ and $\phi_p=\pi/2$. This suppression arises because the transverse momenta $k_y$ in the AM and PM match only within a narrow range as shown in Fig.~\ref{fig:G}(c). In contrast, for other crystallographic orientations such as $\phi_a=\pi/4$ and $\phi_p=0, \pi$, the transverse momenta align more effectively, leading to enhanced conductivity.

Figure~\ref{fig:G}(b) shows that the longitudinal spin conductivity can be either positive or negative, depending on $\phi_a$ and $\phi_p$. In particular, it is negative near $\phi_a = 0$ and becomes positive around $\phi_a = \pi/2$. For $\phi_p \approx 0$ or $\pi$, pronounced (negative) peaks appear in the spin conductivity close to $\phi_a = 0$. As illustrated in Fig.~\ref{fig:G}(d), this behavior originates from the matching of transverse momenta $k_y$ across the junction. The range of matching $k_y$ values between the AM and PM sides is larger for down-spin electrons than for up-spin electrons. Moreover, a greater number of right-moving states on the down-spin Fermi surface (of AM) have velocities close to normal incidence compared to those for up-spins. Consequently, the conductance contribution from down-spin electrons dominates, leading to a negative spin conductivity.

Conversely, for $\phi_a \approx \pi/2$ and $\phi_p = 0$ or $\pi$, the situation reverses—the $k_y$ values of up-spin electrons match over a wider range than those for down-spins, and more up-spin states have velocities near normal incidence. This results in a positive spin conductivity dominated by up-spin transport. Whenever the charge conductivity vanishes, the spin conductivity also vanishes, since no transport occurs. This happens at $\phi_p = \pi/2$ or $3\pi/2$ and $\phi_a = \pi/4$. However, near $\phi_p = 0$ or $\pi$, the charge conductivity remains sizable because both spin species contribute. For $\phi_a = \pi/4$ and all values of $\phi_p$, the spin conductivity is suppressed even though the charge conductivity can be large, as the up- and down-spin contributions nearly cancel, as seen in Fig.~\ref{fig:G}(c).

\begin{figure} [htb]
    \centering
    \includegraphics[width=4.25cm,height=3.7cm]{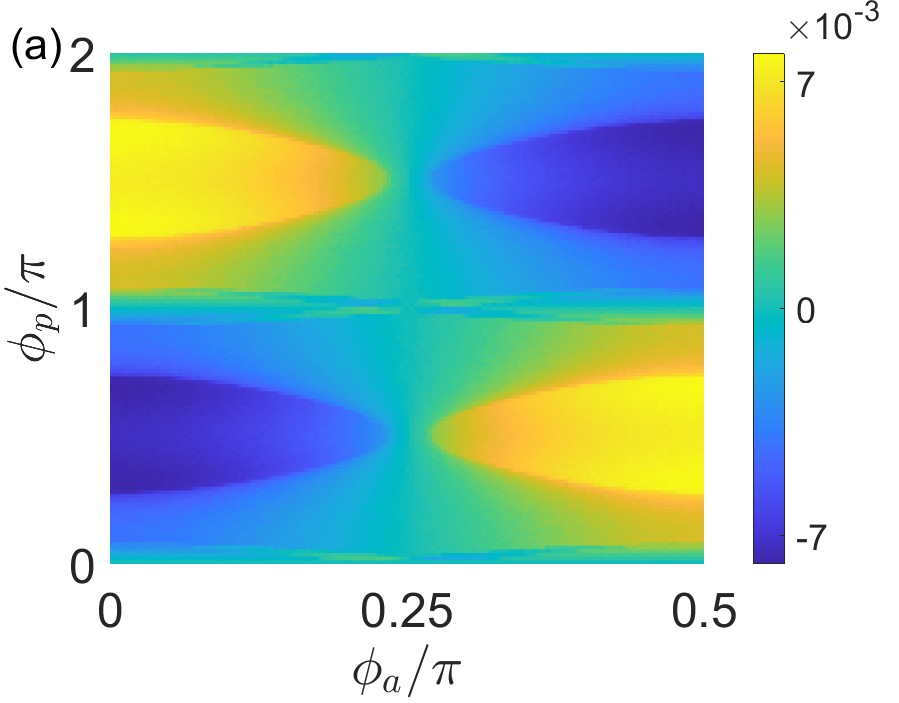}
    \includegraphics[width=4.25cm,height=3.7cm]{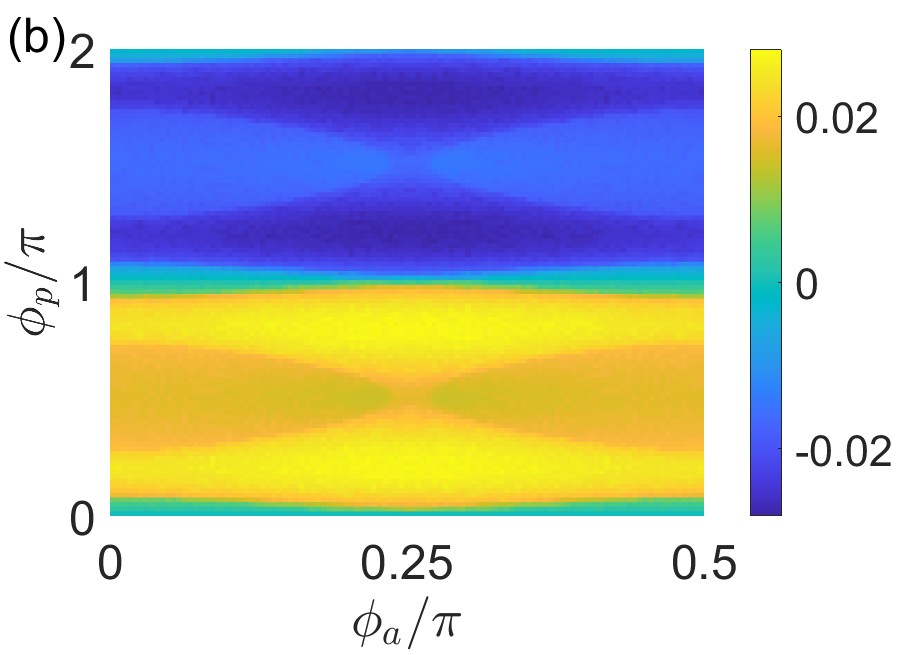} 
    \includegraphics[width=4.25cm,height=3.3cm]{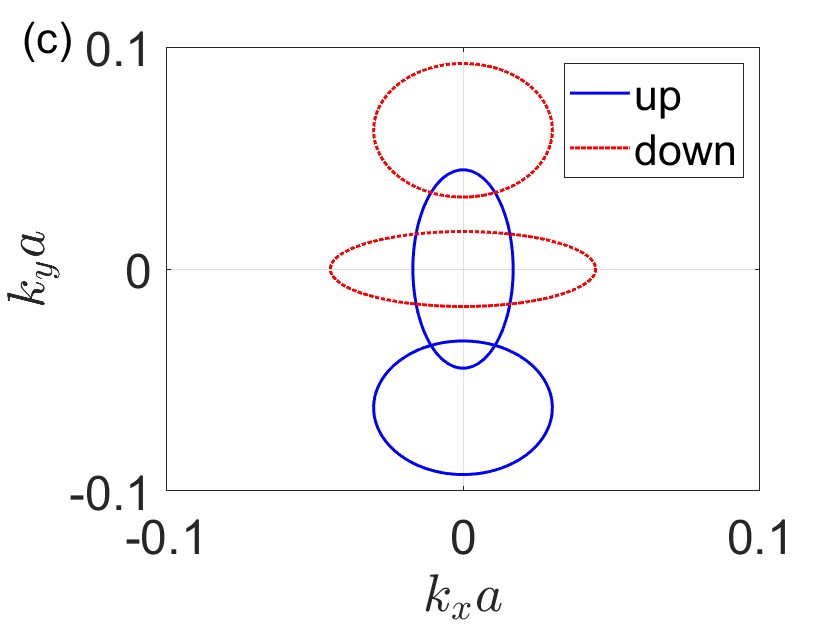}
    \includegraphics[width=4.25cm,height=3.3cm]{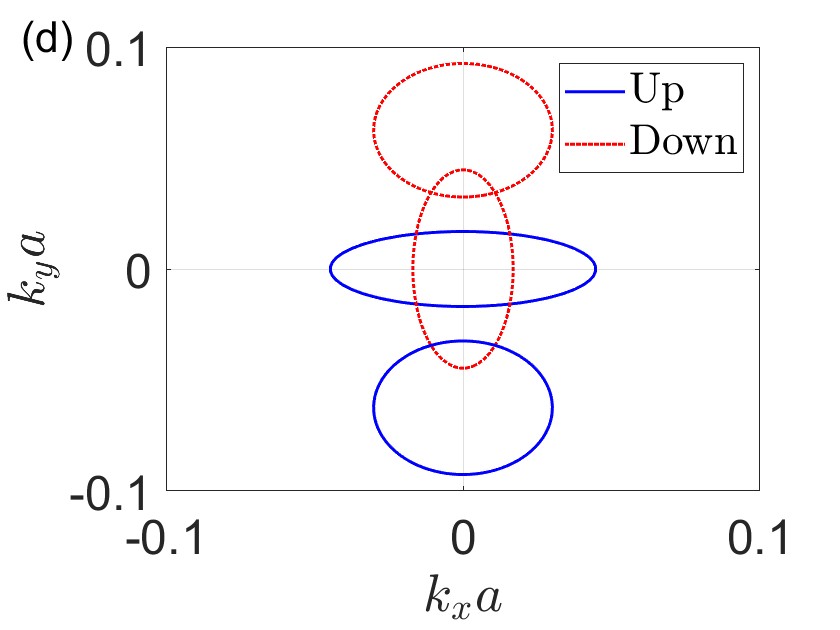} 
        \caption{(a) Transverse charge conductivity in the units of $e^2/ha$, (b) Transverse spin conductivity in units of $e/a$ on the PM at zero bias.  Fermi surface at (c) $\phi_a=\pi/2$ and $\phi_p=\pi/2$ and (d) $\phi_a=0$ and  $\phi_p=\pi/2$ for the same set of parameters as in Fig.~\ref{fig:G}.}
    \label{fig:GT}
\end{figure}

Figure~\ref{fig:GT}(a) shows the transverse charge conductivity as a function of the crystallographic rotation angles of the AM ($\phi_a$) and the PM ($\phi_p$) near the junction at $x=0$, with the same parameters as in Fig.~\ref{fig:G}. For $\phi_a=0$ and $\phi_p=\pi/2$, transverse momentum matching occurs primarily for down-spin electrons. As seen in Fig.~\ref{fig:GT}(d), only a narrow range of $k_y$ values for down-spin electrons matches across the interface, resulting in conduction solely through these modes. Since the contributing down-spin states have negative transverse momentum, the corresponding transverse conductivity is negative. Rotating the AM by $\pi/2$ while keeping $\phi_p=\pi/2$ changes the alignment so that up-spin electrons dominate the transport [Fig.~\ref{fig:GT}(c)], occupying positive transverse momenta and yielding a positive conductivity. Likewise, for $\phi_a=0$ and $\phi_p=3\pi/2$, transport is dominated by up-spin electrons with positive transverse momentum, again producing positive conductivity. In contrast, for $\phi_a=\pi/2$ and $\phi_p=3\pi/2$, conduction is carried mainly by down-spin electrons with negative transverse momentum, leading to negative conductivity.  

Interestingly, for certain $(\phi_a,\phi_p)$ combinations we find that even when the longitudinal charge conductivity vanishes, the transverse conductivity remains finite. This occurs because the available modes in the PM are evanescent: they do not contribute to longitudinal transport but still support a finite transverse current. Moreover, the transverse conductivity is position dependent on the PM side ($x>0$). In particular, when the longitudinal conductivity vanishes, the transverse contribution decays with increasing distance from the junction.  

Figure~\ref{fig:GT}(b) shows the transverse spin conductivity as a function of $\phi_a$ and $\phi_p$ at $x=0$, with the same parameters as before. Near $\phi_a=\pi/4$ and $\phi_p=\pi/2$, the transmitted up-spin electrons occupy positive $k_y$ states while the down-spin electrons occupy negative $k_y$ states. In this case, the transverse charge conductivity nearly cancels due to the opposite contributions, but the transverse spin conductivity peaks because it is given by the difference between the two channels. Similar to the transverse charge conductivity, the transverse spin conductivity also depends on position in the PM region ($x>0$). In particular, for $(\phi_a,\phi_p)$ values where the longitudinal conductivity vanishes, the transverse charge/spin conductivity decays to zero with increasing $x$.

\begin{figure} [htb]
    \centering
    \includegraphics[width=4.3cm]{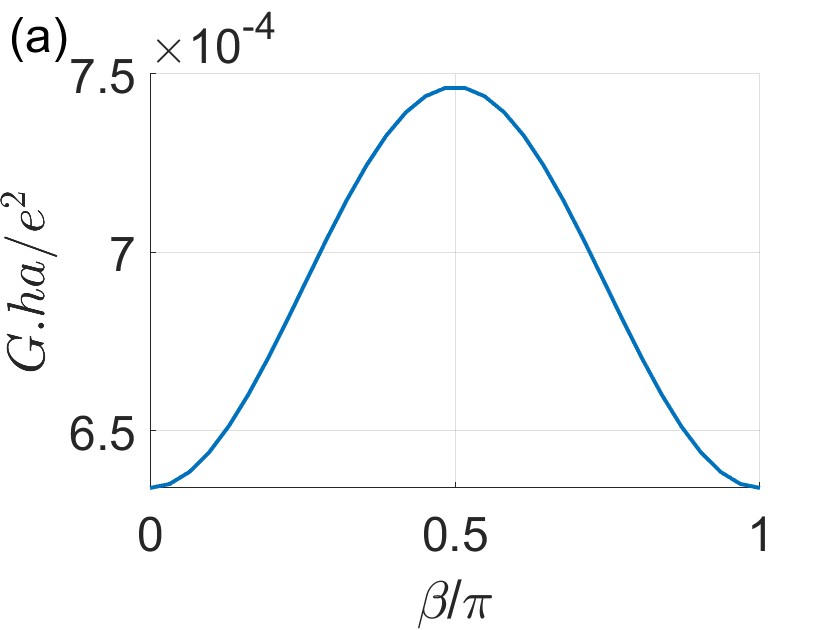}
    \includegraphics[width=4.25cm]{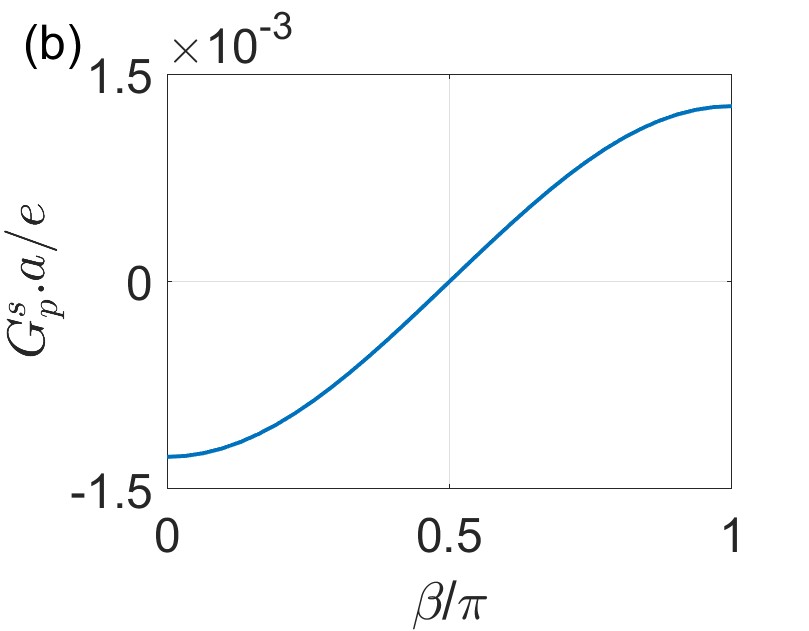}
    \includegraphics[width=4.25cm]{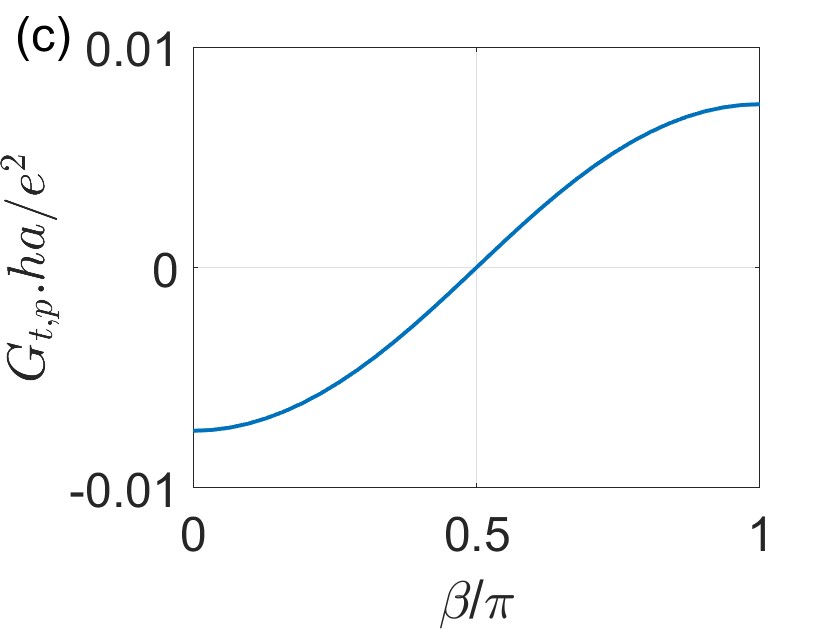}
    \includegraphics[width=4.25cm,height=3.3cm]{revFS_5.jpg} 
    \caption{(a)~Longitudinal charge conductivity in units of $e^2/ha$, (b)~longitudinal spin conductivity in units of $e/a$, (c)~Transverse charge conductivity in units of $e^2/ha$ in the PM, versus  $\beta$ and (d) Fermi surfaces for the same set of  parameters as in Fig.~\ref{fig:G} except for~$\phi_a=0~{\rm and}~\phi_p=\pi/2$.}
    \label{fig:beta}
\end{figure}

Since our system has spin dependent conductivity, rotating the spin quantization axis of PM with respect to that of AM  significantly influences the conductivities on PM as well as AM. In Fig.~\ref{fig:beta}(a) longitudinal charge conductivity in the PM is plotted with respect to $\beta$ for the parameters $\phi_a=0$ and $\phi_p=\pi/2$. For $\beta=0$, the current is carried by the down-spin electrons as can be seen from  fig.~\ref{fig:beta}(d), but as $\be$ changes, the current is carried partially in both the spin channels in the PM, since the up and the down spins on the PM are rotated with respect to those on the AM side. Hence, the variation of longitudinal conductivity versus  $\beta$ is  small which is $\sim 14\%$.

Figure~\ref{fig:beta}(b) shows the longitudinal spin conductivity as a function of $\beta$. For $\beta=0$, the conductivity is negative because transport is dominated by down-spin electrons: only their $k_y$ values match across the junction with the PM. As $\beta$ increases, the orientation of the momentum-matched states gradually rotates from down-spin to up-spin. At $\beta=\pi/2$, the contributions from up- and down-spin electrons are equal and cancel, yielding zero spin current. This cancellation occurs because the modes contributing to transport on the AM side are oriented at $\pi/2$ with respect to the spin quantization axis of the PM.

Figure~\ref{fig:beta}(c) shows the transverse charge conductivity as a function of $\beta$. It is negative for $0<\beta<\pi/2$, increases with $\beta$, crosses zero at $\beta=\pi/2$, and becomes positive for $\beta>\pi/2$. For $\beta=0$, incident down-spin electrons from the AM are transmitted as down-spin states in the PM. Although these states have positive $k_y$, they carry negative group velocity along $\hat{y}$, resulting in negative transverse conductivity. For finite $\beta$, down-spin electrons incident from the AM can be transmitted as either up- or down-spin states in the PM. At $\beta=\pi/2$, equal contributions from up- and down-spin channels with opposite group velocities cancel, producing vanishing transverse conductivity. For $\beta=\pi$, the current is carried entirely by up-spin electrons with positive group velocity, leading to positive transverse conductivity.

\begin{figure} [htb]
    \centering
    \includegraphics[width=4.3cm]{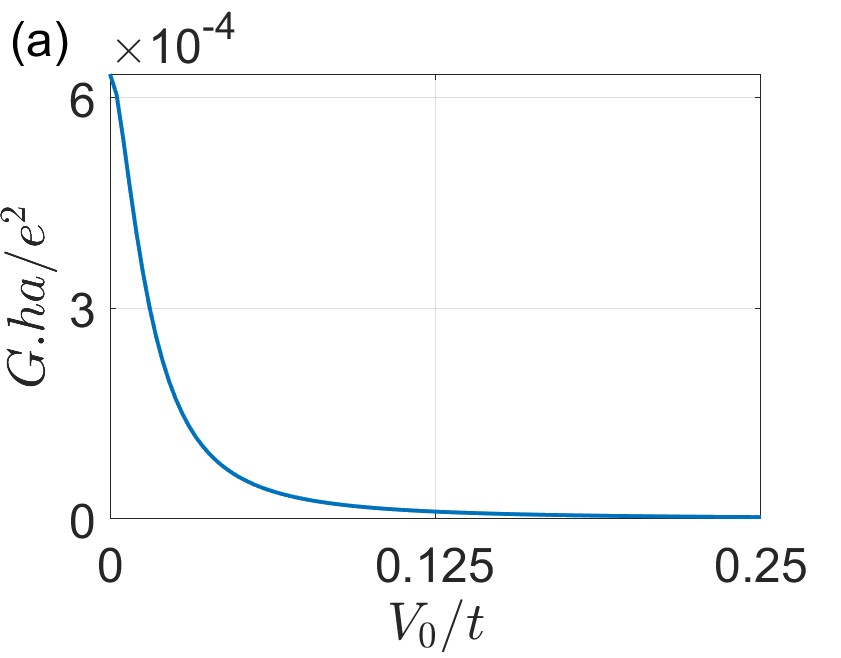}
    \includegraphics[width=4.25cm]{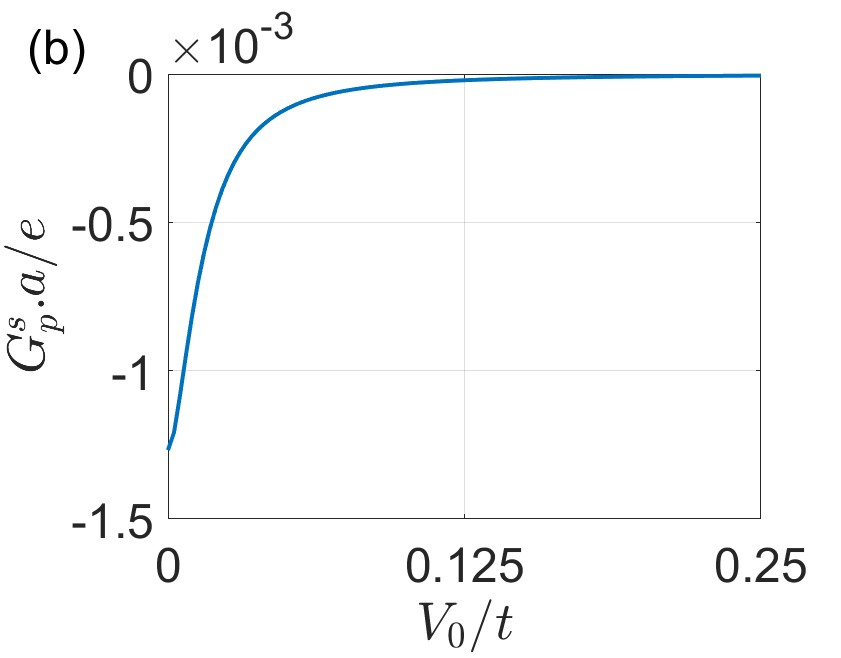}
    \includegraphics[width=4.25cm]{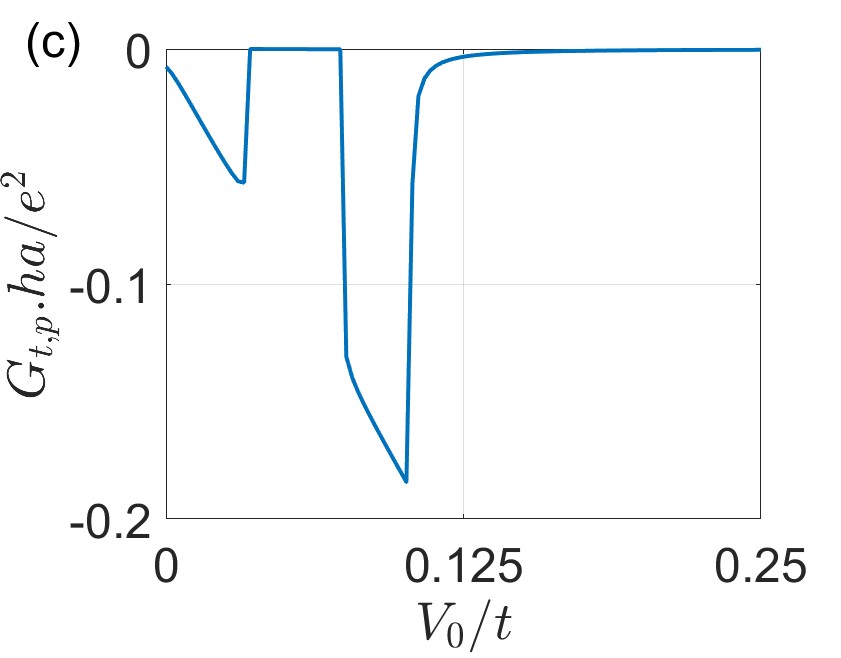}
    \includegraphics[width=4.25cm]{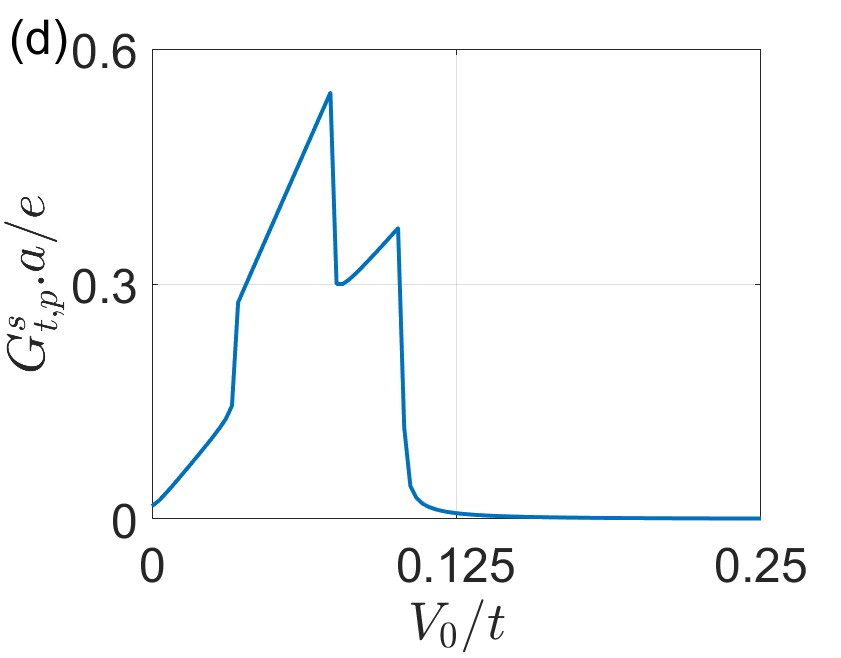}
    \caption{(a), (c)~Longitudinal and transverse charge conductivity in units of $e^2/ha$, (b), (d)~longitudinal and transverse spin conductivity in units of $e/a$ versus $V_0$,  for the same set of  parameters as in Fig.~\ref{fig:G} except for~$\phi_a=0~{\rm and}~\phi_p=\pi/2$.}
    \label{fig:V0PM}
\end{figure}

Figure~\ref{fig:V0PM} shows the dependence of the conductivities on the barrier strength $V_0$ for $\beta = 0$, $\phi_a = 0$, and $\phi_p = \pi/2$, evaluated on the PM side of the junction. The longitudinal charge and spin conductivities are identical on both sides of the junction and decay monotonically to zero with increasing $V_0$, similar to the transmission behavior across a potential barrier in a normal metal. In contrast, the transverse conductivities display a nonmonotonic dependence on $V_0$ at small barrier strengths. This behavior is in quantitative agreement with the transverse charge and spin conductivities obtained from the analytical expressions for the scattering coefficients in Eqs.~\eqref{eq:V0_up} and \eqref{eq:V0_down}.

\subsection{In  altermagnet}~\label{sec:AM-result}
\begin{figure} [htb]
    \centering
    \includegraphics[width=4.25cm,height=3.7cm]{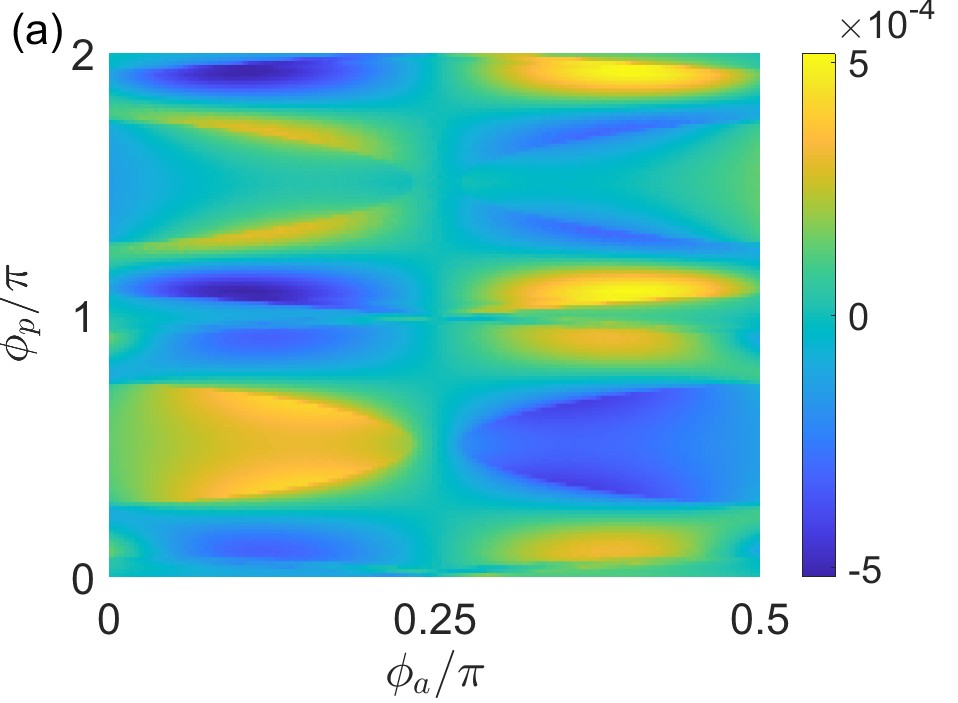}
    \includegraphics[width=4.25cm,height=3.7cm]{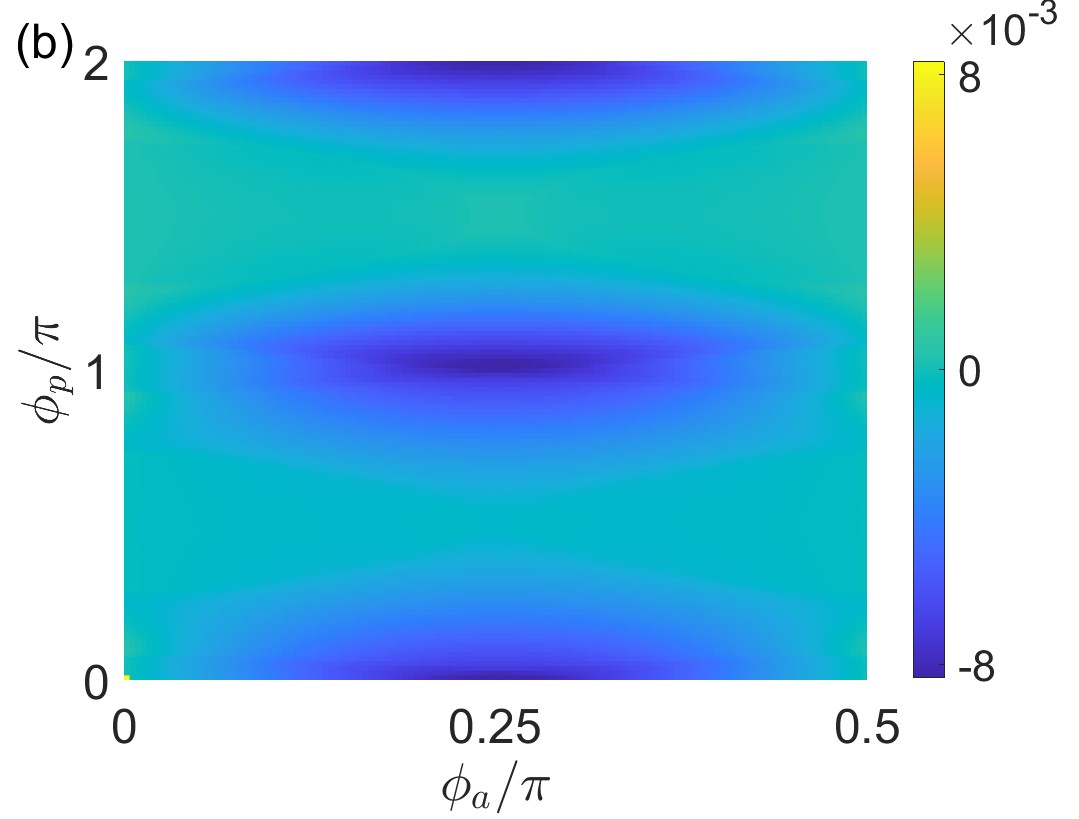}
    \caption{(a) Transverse charge conductivity in units of $e^2/ha$ (b) Transverse spin conductivity in units of $e/a$ in the AM  with respect to $\phi_a$ and $\phi_p$ at $x=0$  and zero bias. Other parameters used are same as Fig.~\ref{fig:G}.}
    \label{fig:AM1}
\end{figure}

Figure~\ref{fig:AM1}(a) shows the transverse charge conductivity in the AM at $x=0$ as a function of $\phi_a$ and $\phi_p$. Peaks in the conductivity appear in parameter regions where the longitudinal charge conductivity vanishes. This indicates that even perfectly reflecting modes contribute to the transverse charge current. From Eq.~\eqref{eq:Jyam}, one might expect the first two terms to vanish upon summation over all incident angles, since they have a multiplicative factor of $k_y$. However, these terms also contain $\psi^{\dagger}\psi$ and $\psi^{\dagger}\sigma_z \psi$, which are not exactly equal for angles of incidence  $\theta$ and $-\theta$, because the $k_y \to -k_y$ symmetry is broken for each spin on the PM Fermi surface. Moreover, $\psi^{\dagger}\psi$ and $\psi^{\dagger}\sigma_z\psi$ depend not only on the reflection amplitude but also on its phase, which differs for $\theta$ and $-\theta$. These phases become equal when the PM is aligned at $\phi_p=0$ or $\pi$, and in this case the transverse conductivity vanishes. Since the breaking of $k_y \to -k_y$ symmetry is maximal for $\phi_p=\pi/2$ and $3\pi/2$, the transverse conductivity peaks near these orientations for fixed $\phi_a$. Similarly, for $\phi_a=\pi/4$, the transverse charge conductivity is zero for all $\phi_p$, because the current is carried equally by both spin species whose $k_y$ contributions cancel. 

Figure~\ref{fig:AM1}(b) shows the transverse spin conductivity as a function of $\phi_a$ and $\phi_p$.  For $\phi_a=\pi/4$, the transverse spin conductivity remains finite even though the charge conductivity vanishes. This occurs because the transverse charge conductivity involves the sum of contributions from all incidence angles $\theta$, which cancel between $\theta$ and $-\theta$ due to opposite transverse velocities. In contrast, the spin conductivity involves the difference between spin channels and therefore does not vanish under the same symmetry.

\begin{figure} [htb]
    \centering
    \includegraphics[width=4.25cm,height=3.8cm]{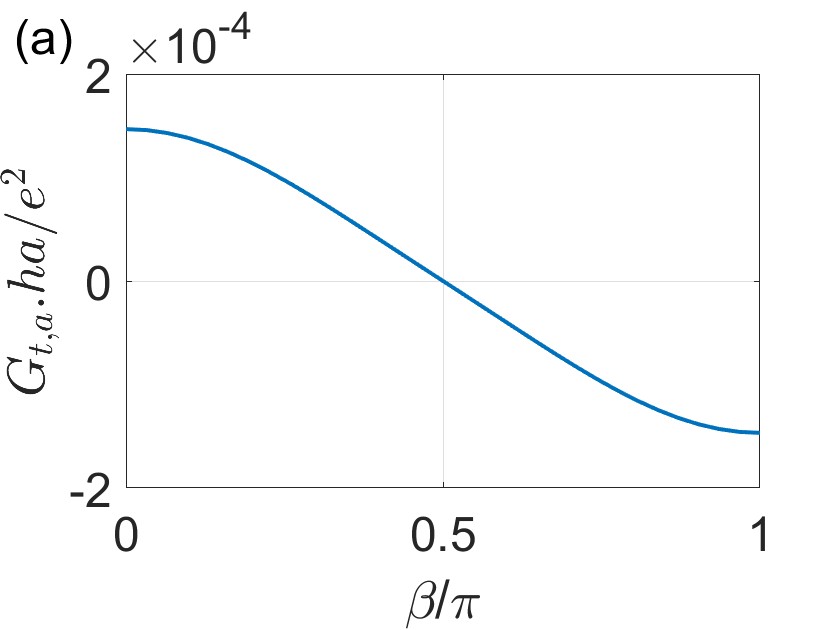} 
    \includegraphics[width=4.25cm, height=3.8cm]{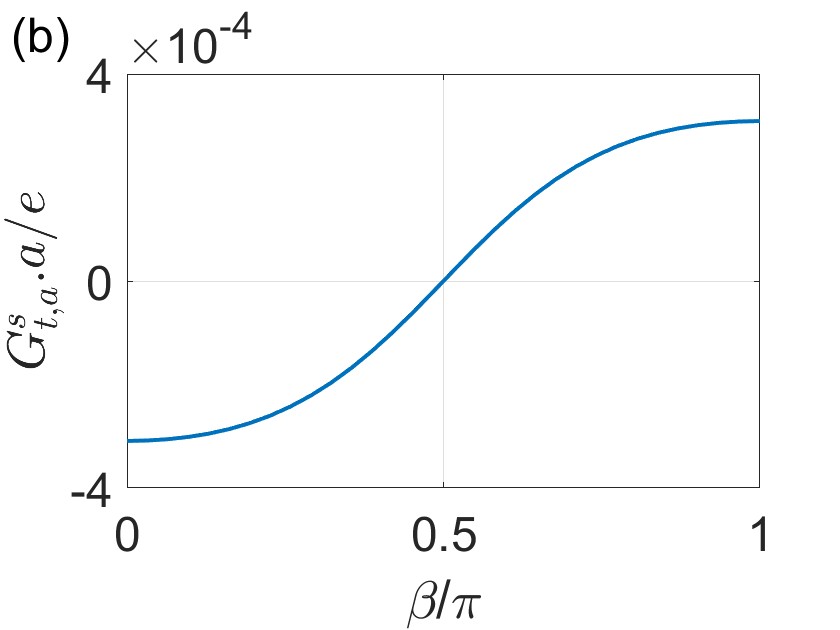}
    \caption{(a) Transverse charge conductivity in units of $e^2/ha$ (b) Transverse spin conductivity in units of $e/a$ in the AM versus $\beta$ at zero bias. Other parameters used are same as in Fig.~\ref{fig:beta}.  }
    \label{fig:betaAM}
\end{figure}

Figure~\ref{fig:betaAM}(a) shows the variation of transverse charge conductivity, while Fig.~\ref{fig:betaAM}(b) depicts the corresponding spin conductivity as a function of $\beta$. For $\beta < \pi/2$, the transverse charge conductivity is positive. This can be understood by noting that for  $\phi_a=0$ and $\phi_p=\pi/2$,   transmission occurs predominantly for down-spin electrons with positive $k_y$ in a narrow range (see Fig.~\ref{fig:beta}(d)). On the AM side, reflection takes place at negative $k_y$, but these states carry positive velocity, leading to a net positive transverse current. At $\beta=\pi/2$, reflection occurs symmetrically for positive and negative $k_y$, and the associated velocities cancel, resulting in vanishing transverse charge conductivity. For $\pi/2 < \beta < \pi$, the situation is reversed, giving rise to negative transverse charge conductivity. The current on the AM side is predominantly carried by down-spin electrons, leading to transverse charge and spin conductivities of opposite sign. Consequently, as shown in Fig.~\ref{fig:betaAM}(b), the transverse spin conductivity exhibits a trend opposite to that of the transverse charge conductivity in Fig.~\ref{fig:betaAM}(a).

\begin{figure} [htb]
    \centering
    \includegraphics[width=4.25cm]{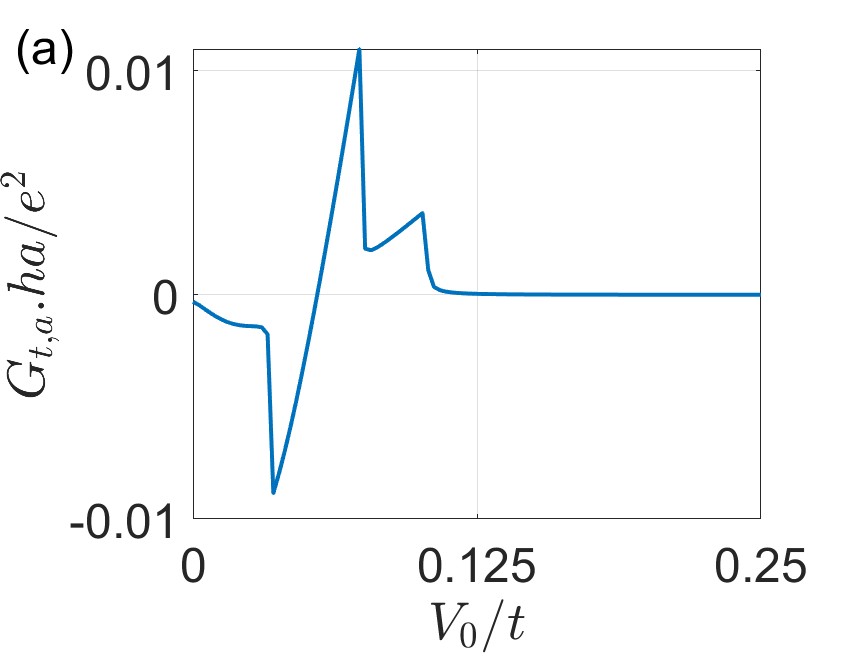} 
    \includegraphics[width=4.25cm]{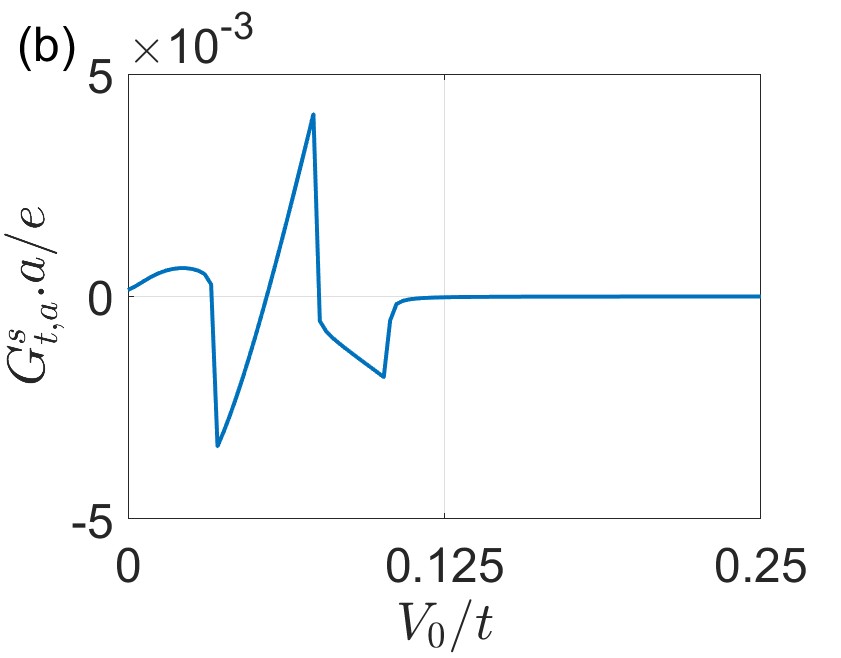}
    \caption{(a)~Transverse charge conductivity in units of $e^2/ha$, (b)~Transverse spin conductivity in units of $e/a$ versus $V_0$,  for the same set of  parameters as in Fig.~\ref{fig:G} except for~$\phi_a=0~{\rm and}~\phi_p=\pi/2$.}
    \label{fig:V0AM}
\end{figure}
Figure~\ref{fig:V0AM} shows the dependence of the transverse charge and spin conductivities on the barrier strength $V_0$, evaluated on the AM side of the junction. The conductivities exhibit a nonmonotonic variation with $V_0$, which arises because they are evaluated in the immediate vicinity of the interface at $x = 0^-$. This behavior can be understood from the analytical expressions for the scattering coefficients given in Eqs.~\eqref{eq:V0_up} and \eqref{eq:V0_down}, which capture the detailed dependence of the transverse transport on the barrier potential.

\section{Discussion}~\label{sec:disc}

The transverse charge conductivities obtained in our calculations are expressed in units of $e^2/ha$, which corresponds to approximately $4000~\Omega^{-1}\mathrm{cm}^{-1}$. In physical units, the computed transverse charge conductivities are of the order of $1$–$10~\Omega^{-1}\mathrm{cm}^{-1}$. For comparison, anomalous Hall conductivities ranging from $20$ to $600~\Omega^{-1}\mathrm{cm}^{-1}$ have been reported in various antiferromagnets~\cite{Yamada25}. The transverse spin conductivities in our system are of the order of $480~(\hbar/e)~\Omega^{-1}\mathrm{cm}^{-1}$ in the $p$-wave magnet and $180~(\hbar/e)~\Omega^{-1}\mathrm{cm}^{-1}$ in the altermagnet. These values are substantial when compared with the reported spin Hall conductivity of $1710~(\hbar/e)~\Omega^{-1}\mathrm{cm}^{-1}$ in tungsten~\cite{Ishikawa23} and $0.18~(\hbar/e)~\Omega^{-1}\mathrm{cm}^{-1}$ in the semiconductor ZnSe~\cite{Stern06}. Hence, the magnitude of the transverse spin conductivity we predict is quite significant.

Importantly, we have demonstrated that both transverse charge and spin conductivities can arise in altermagnet-$p$-wave magnet junctions even in the absence of spin-orbit coupling, in contrast to conventional spin Hall effects that rely on spin-orbit interaction.

Within the scattering framework, an alternative approach would be to use a lattice model for the AM and PM with spin-dependent nearest-neighbor hopping. However, such lattice models are limited in that they do not allow the crystallographic orientation of the AM to be rotated by arbitrary angles while preserving translational invariance along the $y$ direction. In contrast, the continuum model employed in this work treats $\phi_a$ as a tunable parameter in the Hamiltonian, enabling systematic exploration of arbitrary crystallographic rotations.

\section{Summary and Conclusions}~\label{sec:summary}

We have investigated charge and spin transport across junctions of AMs and PMs within a continuum model, incorporating arbitrary crystallographic orientations and relative spin quantization axes. By framing the boundary conditions and enforcing transverse momentum matching, we obtained the longitudinal and transverse conductivities on both sides of the junction.

Our analysis shows that despite both AM and PM being spin-unpolarized materials, the junction supports finite spin currents in both longitudinal and transverse channels. The longitudinal charge conductivity is conserved across the junction, whereas the longitudinal spin current is generally not conserved because of the absence of a globally commuting spin operator. Interestingly, we find parameter regimes where the longitudinal charge conductivity vanishes but the transverse charge and spin conductivities remain finite, highlighting the unconventional role of mode matching and spin splitting in these systems.

The interplay between the crystallographic orientations ($\phi_a$, $\phi_p$) and the relative spin quantization axis $\beta$ governs the transport response. Depending on the relative alignment, conduction is dominated by either up- or down-spin modes, giving rise to sign changes in both charge and spin conductivities. In particular, transverse charge and spin conductivities often exhibit opposite signs, with their magnitudes controlled by mismatches in $k_y$ across the interface. For certain orientations, the transverse spin conductivity survives even when the transverse charge conductivity cancels, underscoring their distinct microscopic origins.

These results establish that AM–PM junctions can host anomalous Hall and spin Hall effects without requiring spin–orbit coupling. Moreover, we demonstrate that transverse responses can outweigh the longitudinal ones for suitable orientations and parameter regimes, offering a high degree of tunability. Taken together, our findings position AM–PM junctions as a promising platform for realizing unconventional spin currents and controllable Hall responses, thereby broadening the scope of spintronics beyond conventional ferromagnets and spin–orbit-coupled systems.

 Our predictions can be tested in realistic material platforms. Prominent candidates for AMs include KRu$_4$O$_8$, Mn$_5$Si$_3$, and KV$_2$Se$_2$O, while PMs such as NiI$_2$, CeNiAsO, and Mn$_3$GaN provide partners to form heterojunctions\cite{Tomas22,Helena21,Jiang2025,Hellenes24,Song2025}. The results obtained are for parameters of KRu$_4$O$_8$ and CeNiAsO.  Junctions fabricated from these materials would allow direct measurements of the anomalous Hall and spin Hall effects in the absence of spin–orbit coupling, as well as the spin currents predicted on both sides of the interface.  Thus, these material realizations place our theoretical predictions within experimental reach, strengthening the case for AM–PM junctions as versatile platforms for spintronics.
\begin{acknowledgments}
We thank Bijay Kumar Sahoo for useful discussions. We thank Science and Engineering Research Board (now Anusandhan National Research Foundation) - Core Research grant (CRG/2022/004311) for financial support. 
\end{acknowledgments}
\bibliography{alt_pwave}
\end{document}